\documentclass[useAMS,usenatbib,usegraphicx]{mn2e}

\def\kms{$\rm km\, s^{-1}$}
\def\cm3{$\rm cm^{-3}$}
\def\Ts{$\rm T_{*}$}
\def\Vs{$\rm V_{s}$}
\def\n0{$\rm n_{0}$}
\def\B0{$\rm B_{0}$}

\def\Fh{$\rm F_{H}$}
\def\Hb{H$\beta$}
\def\Ha{H$\alpha$}

\def\Fh{$\rm F_{h}$}
\def\erg{$\rm erg\, cm^{-2}\, s^{-1}$}

\def\L12{L$_{12\mu m}$~}
\def\F12{F$_{12\mu m}$~}

\title[The structure of the LLAGN NGC 4579]{The complex structure of  low luminosity active galactic nuclei : NGC 4579}
\author[M. Contini]{Marcella Contini \thanks{E-mail:
contini@post.tau.ac.il} \\
School of Physics and Astronomy, Tel Aviv University, Tel Aviv 69978, Israel\\}
\begin{document}

\date{Accepted . Received ; in original form }

\pagerange{\pageref{firstpage}--\pageref{lastpage}} \pubyear{2002}

\maketitle

\label{firstpage}

\begin{abstract}

We have modelled the  low luminosity AGN NGC 4579
by  explaining both the continuum and the line spectra  observed  with different apertures.
It was found that the nuclear emission is  dominated by an AGN
such that the  flux from the active centre (AC) is relatively low compared with that 
of  the NLR of Seyfert galaxies. However, the contribution of a young starburst
cannot be neglected, as well as that of  shock dominated clouds with velocities of 100, 300, and 500 \kms.
A small contribution from an older starburst with an age of 4.5 Myr, probably
located in the external nuclear region is also found.
 HII regions appear in the
 the extended  regions ($\sim$ 1 kpc),  where radiation  and shock dominated clouds  with \Vs=100 \kms
prevail. 
The continuum SED of NGC 4579 is characterized by
the strong flux from  an old stellar population.
Emissions in the radio range show 
synchrotron radiation from the base of the jet outflowing from the accretion disc 
 within 0.1 pc from the active centre.
Radio emission  within intermediate distances (10-20 pc) is explained by the bremsstrahlung
from gas downstream of low velocity  shocks (\Vs=100 \kms) reached by a  relatively  low
radiation flux from the AC. 
In  extended regions  ($>$ 100 pc) the radio emission is synchrotron
radiation  created by the Fermi mechanism at the shock front. 
The shocks are created by  collision of clouds with the jet.
All types of emissions observed at different  radius from the centre
can be  reconciled   with the presence of the  jet.

\end{abstract}


\section{Introduction}

The analysis of the line and continuum spectra of active galactic nuclei
(AGN) developed in the last years,  revealed that radiation  
in the different spectral ranges 
depends on  different  excitation mechanisms of  gas and dust.
It was found, in particular  (e.g. Contini \& Viegas 2000, Contini, Viegas \& Prieto 2004) 
that both the photoionizing flux from the active centre (AC) and collisional processes
in supersonic velocity fields   contribute to  ionization and heating
of the emitting gas (and dust) in the  narrow  emission-line region (NLR) with different relative importance.

In low luminosity objects, such as LLAGNs (low luminosity AGN) and LINERs
(low ionization nuclear emission  regions) 
starbursts as well as an active nucleus   can be present with  comparable importance. 
Anyway,  winds and jets  are often observed,  leading to a prominent role of shocks.
The complex nature of LINERs  was  revealed by modelling  the most  
significant line ratios in the UV-optical-IR spectra (Contini 1997).
Moreover, modelling the X-ray -IR continuum correlation in a large sample of AGN,
Contini, Viegas \& Campos (2003) found that LLAGNs are in the low luminosity tail 
of AGN. This is  explained by  a relative low  flux from the AC,  roughly two orders of
magnitude lower than  that found in the NLR of Seyfert 2 and Seyfert 1 galaxies,

In this paper we   analyse the spectra of  NGC 4579  which  is classified
 as a LINER 1.9 galaxy with many AGN characteristics by Pogge et al. (2000)
and as a Seyfert 1.9  by Ulvestad \& Ho (2001).
The morphological type of NGC 4579 is SBb. The galaxy is located  at a distance of 20 Mpc (cz=1502 \kms) (Gonzalez-Delgado \& Perez, 1996).
Hubble Space Telescope (HST) images at 2300 \AA ~(Maoz et al. 1995) show emission resolved into two
components, a bright core and a weaker source  with  a flux corresponding to
15 \% of that of the bright  core. The weaker source is
extended 0.1 X 0.2 arcsec$^2$ and is located  0.2 arcsec from the bright  core at PA=72$^o$.

From ground-based observations the galaxy was previously known to contain a type 1 nucleus
revealed by a broad \Ha ~ line  (FWHM of $\sim$ 2300 \kms). 
New spectra obtained by HST  0.2"  slit
reveal an \Ha ~ component with FWZI of $\sim$ 18,000 \kms (Barth et al, 2001)
with  "shoulders" on the red and blue sides.
About 31 low-luminosity HII regions have been detected and measured in the disc, located
along the dust lanes of the spiral arms, at distances between 3.3 and 7.5 kpc from the nucleus
(Gonzalez-Delgado \& Perez).

 NGC 4579 line spectra have been  observed  with different apertures in the different frequency
 domains by
Gonzalez-Delgado \& Perez (1996) within 1.5" in the nuclear region and within 20" in regions A and B,
and previously by Keel (1983) within 4.7", and by Stauffer (1982) within 4.0".

In this paper we  address the  line spectra observed by Gonzalez-Delgado \& Perez 
 with small and large apertures. The latter include  
 some information   about the surroundings  of the nucleus.
The  earlier data by Keel and Stauffer obtained with intermediate apertures 
will be considered 
only to check the consistency of  modelling.
In fact, the large apertures  integrate on all  the conditions found in the 
nucleus,  the values from the small aperture and the surroundings.
In the case of this galaxy they are rather complex and includes every thing 
from a nuclear stellar bar, a starforming ring, an the old galaxy 
population, etc. 

Indeed, the low [OIII] 5007/4363 observed line  ratio (Sect. 2.2) indicates relatively high temperatures  
in the emitting gas, because
 very high densities ($\geq$ 1000 \cm3)  downstream are excluded by  [SII] 6717/6730 $\geq$ 1. 
The radiation flux
from the AC cannot heat the gas of the NLR to temperatures $>$ 2 10$^4$ K, which, on the other hand,
 are easily reached by collisional heating downstream of the shock front.
Moreover, the [OII] 3727+/\Hb
~(the + symbol indicates that the
doublet intensities are  summed up) and  [OI] 6300+/\Hb ~line ratios are relatively high.
So  pure photoionization models (cfr. the code CLOUDY) are less  adapted to  model NGC 4579.
The  SUMA code (Viegas \& Contini 1994, Contini \& Viegas 2001a,b
and references therein) which  calculates the spectra emitted  from gas and dust 
accounting  for the coupled effects of shocks and  of  an external photoionizing source,  is adopted
throughout this work.

The  spectral energy distribution (SED) of the continuum   is  calculated  consistently 
with the line spectra to constrain the models. 
Particularly, the SED of  NGC 4579 continuum
is characterised by  an unusually high contribution 
 of  the old stellar population, higher by a factor of $\sim$ 10 than in Seyfert 2
  and Seyfert 1 galaxies.

The observations in the radio  by Anderson, Ulvestad, \& Ho (2003) provide the
fluxes within 10$^4$ Schwarzchild radii of the black hole, i.e. within 0.1 pc from the nucleus
for a sample of LLAGN
(NGC 3147, NGC 4168, NGC 4203, NGC 4235, and NGC 4450).
NGC 4579   was actually selected from this sample  because  radio data 
are available also  on larger scales.
It was suggested by Ulvestad \& Ho (2001) that jets are responsible of  
the subparsec radio emission.  This  may originate in synchrotron radiation from discrete plasma
components or from the "base" of a continuous jet ejected from the central engine, which becomes
optically thin on larger scales (Nagar, Wilson, \& Falcke 2001).
Anderson et al.
explain the observed spectral slope by {\it compact jets that can accomodate the average
spectral index, the relatively high radio luminosity, and the unresolved appearance.}
This applies only
if the jets are fairly close to our line of sight.  Anderson et al.
 conclude that, as the galaxies show a
broad line component, they have to be in the right direction.
This suggests that synchrotron emission with different spectral  indices can be observed at
different distances from the AC, and that shock fronts are at work.
The shocks are created  by collision  of jets with the  interstellar  medium (ISM).
The in-situ particle acceleration associated with shocks could lead then to extended regions
of radio emission with suitable spectral indices.
Moreover,  if the jets were  disrupted
by interaction with the ISM, the emission line and continuum spectra from the NLR  clouds  should 
show the characteristic features of shocks, such as  relatively high low ionization (e.g. [OII],[NII], 
and [SII]) and neutral (e.g. [OI]) line ratios to \Hb, and high [OIII] 4363 /[OIII] 5007+.
Moreover, soft X-ray emission and dust reradiation in the IR could be explained,
as well as  faint diffuse radio emission (Middleberg et al. 2004).
The signature of shocks is well recognizable  in the radio range 
of the continuum   because the power-law type  of synchrotron radiation
has a   different slope   than that of bremsstrahlung from cool  gas (Contini \& Viegas 2000).

In this paper, we will investigate whether  radio emissions   within different radii 
from the AC on large scales are consistent with the jet hypothesis by
 showing that the different types of emissions  within different
regions  can be   explained  by the shocks.
Actually, shocks in the NLR of AGN are  investigated by modelling
the line and continuum spectra. 
In order to unravel   the different  aspects of  the NGC 4579
complex  (e.g. the AGN, the starburst, and the HII region), 
we will model NGC 4579  first through the line spectra  (Sect. 2).
We will calculate the different components of  the continuum SED and compare them
with the data observed with  different  apertures in Sect. 3. Discussion and conclusions
follow in Sect. 4.

\section{The line spectra}

 The spectra in the nuclear region
observed
 by Gonzalez-Delgado \& Perez (1996)
within 1.5" at  the 4.2 m William Herschel Telescope in La Palma during 1988 March,
and at   the 1 m Jacobus Kapteyn Telescope during 1992 May
show  a  number of line ratios large enough for modelling.
The spectra observed in the circumnuclear region ($\leq$ 20") contain a small number of lines,
and are not reddening  corrected. They can however provide some precious results.
The line spectra observed by  Keel (1983) at 4.7" and  by Stauffer (1982) at 4.0"
at the Lick 1 m Anna Nickel telescope are relatively poor in number of lines.

 Pogge et al (2000) in their Table 3 give  the ratio of [OIII]/\Ha ~in the nuclear (0.23", 18pc)
region and in the
circumnuclear (2.5", 194pc) regions, which correspond to  [OIII]/\Ha=0.37 and 0.06 , respectively.
However, they cannot be used for the modelling because they refer to "band" fluxes, not converted
 to emission in a particular line by correcting for the filter transmission
of other lines in the bandpass.

\subsection{Modelling procedure}

We consider
clouds moving outwards from the centre. The inner edge of the clouds is illuminated  by the  flux from
the  photoionizing source (AGN, starburst, HII region, etc) while the outer edge
defines  the shock front.

 As a first guess, we compare the data with   models
from the grids calculated by the SUMA code (Contini \& Viegas 2001a,b, thereafter CV01a and CV01b).
The models presented in (CV01a) correspond to the AGN with power-law flux intensities ranging from
zero in  shock dominated (SD) clouds up to a maximum of log \Fh $\sim$ 13
which was found in radiation dominated (RD) clouds by previous modelling of AGN
(e.g. Contini \& Viegas 2000).
  Models presented in CV01b  refer to the starbursts where the ionizing flux corresponds to
a stellar cluster. Models for HII regions  where  black body radiation
 reaches the inner edge of the cloud are also presented in CV01b.
All the models  are composite, i.e. they account also for the shocks,
as suggested by the  relatively strong low ionization level lines
(Contini 1997),  by soft X-ray,  as well as by infrared luminosities
(Contini, Viegas, \& Campos 2003).

The input parameters  of SUMA  relative to the shock are the following :
 the shock velocity, \Vs,  the pre-shock density, \n0 and the pre-shock magnetic field, \B0.
Those relative to the flux are
  the power law flux, \Fh, in units of photons cm$^{-2}$ s$^{-1}$ eV$^{-1}$ at 1 Ryd,
the spectral indices in the UV and X-ray ranges. $\alpha_{UV}$ = -1.5 and
$\alpha_{X}$ = -0.4 are adopted, respectively.
In models representing starbursts,
t is the starburst age, and for HII regions \Ts ~is the  color temperature of the stars.
U is the ionization parameter in both cases.
For all types of models
D is the geometrical thickness of the cloud, and d/g the dust-to-gas ratio.
A preshock magnetic field  \B0 = 10$^{-4}$ gauss  and cosmic relative abundances (Allen 1973)
are used for all the models.

The intensity of each line depends on the physical conditions  of the emitting gas, namely,  on the
distribution of the temperature, density, and fractional abundance of the corresponding ion downstream
of the shock front,  and on the relative abundance of the elements.

In  our previous  modelling of the NLR of AGN (e.g. Contini et al. 2003 and references therein)
it was found that  multi-cloud models  explain the observed spectra.
These models
result  from the weighted sum of models representing  single clouds  
each characterized mainly by
 the shock velocity, the gas density,  and the  radiation flux intensity.
 For LINERs, in particular, it was found that the single-cloud models
 account for  all types of ionizing
and heating mechanisms (Contini 1997), e.g. the power-law radiation from the active nucleus,
the radiation from  starburts, the black body radiation from
stars with a certain colour temperature, and shocks created by collision of clouds
in the ISM and in  supernova remnants (SNR).
Therefore multi-cloud models will be adopted in the present modelling.
The relative  importance of the different types  which best fit the observed spectra
will  shed some light on  the characteristics of the galaxy on scales larger than the nuclear ones.

\begin{figure}
\includegraphics[width=88mm]{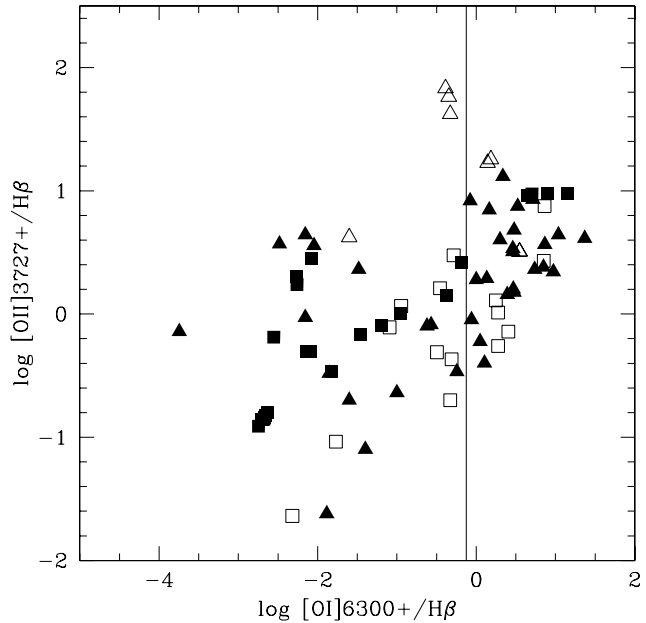} 
\caption
{[OII]/\Hb ~versus [OI]/\Hb ~calculated by the grids. Black triangles represent models
calculated by a power-law radiation flux; open triangles refer to shock dominated models;
open squares represent models calculated by black body, and black squares refer to starburst.
The vertical solid line indicates [OI]/\Hb = 0.75.
}
\end{figure}

\begin{table*}
\centering
\caption{Single-cloud models}
\begin{tabular}{lllllllllllllll}\\ \hline
\             & M1 & M2 & M3 & M4 & M5 & M6 & M7 & M8 & M9 & M10 & M11 & M12       \\
\ type$^1$    & SD & SD & SD & pl & pl & pl & pl & pl & sb & sb  & sb  & hII     \\
\hline
\ [OII]  3727+   & 4.20 & 16.80  & 3.20  & 4.0  & 4.80  &13.10  & 8.30  & 7.0  & 2.60  & 1.73 &  9.50 &  1.0\\
\ [NeIII]  3869+ &  .17 &  2.70  & 1.0  &  .50  &  .94  &  .77  &  .86  & 1.21  &  .33  &  .54 &  3.45 &   .12\\
\ [OIII]  4363   &  .12 &  1.35  &  .31  &  .02  &  .03  &  .29  &  .04  &  .03  &  .01  &  .05 &   .07 &   .19\\
\ [OIII]  5007+  & 1.55 & 17.10  & 4.60  & 1.70  & 7.40  & 3.67  & 3.86  & 6.42  & 2.20  & 9.54 & 13.17 &  2.0\\
\ [NI]  5200+    &  .01 &   .26  &  .53  &  .18  &  .17  &  .04  &  .003  &  .03  & 0.03 & 3.(-4)& .025&    .13\\
\ [OI]  6300+    &  .03 &  1.39  & 3.54  & 2.0  & 3.0  & 2.17  &  .84  & 1.46  &  .66  &  .005 &  8.04 &  1.90\\
\ [NII] 6548+    &  .46 &  4.50  & 3.34  & 4.90  & 7.43  & 7.81  & 5.0  & 5.79  & 2.64  &  .88 & 13.94 &  5.0\\
\ [SII] 6716     &  .15 &  1.50  & 2.60  & 3.86  & 5.20  & 3.73  & 2.21  & 1.65  & 1.80  &  .14 &  1.93 &  1.75\\
\ [SII] 6730     &  .14 &  2.27  & 4.06  & 3.12  & 4.34  & 3.20  & 2.19  & 2.43  & 1.70  &  .15 &  1.71 &  1.95\\
\ [OII] 7325+    &  .34 &  5.15  &  .79  &  .12  &  .13  &  .79  &  .25  &  .39  &  .06  &  .04 &   .25 &   .03\\
\ [SIII] 6069+   &  .14 &  1.40  &  .24  & 1.38  & 1.77  & 1.28  & 1.17  & 1.60  &  .24  & 2.65 &  9.13 &  3.80\\
\ H$\beta^2$     & .81  &  .75  & 6.7   &6.   & 12.6    & .35   &3.   & 45.    & 16.    & 5.80 & 68.  & 700.  \\
\ \Vs ~(\kms)   & 100  &  300  &  500 &   100  &  100   & 100   & 100  &  100  &  100   & 100 &   100 &   200   \\
\ \n0 (\cm3)     & 100  &  300  &  300 &   100  &  100   & 100   & 100  &  300  &  100   & 100 &   100 &   200   \\
\ D (10$^{17}$cm)& 100  &  100  &   10 &   100 &  100   &   1   &   1  &    1  &  100   &   1 &   100 &   100   \\
\ log F$_h^3$    &  -   &  -    &   -  &  9.    & 9.48   &8.     &9.    &10.    &  -     & -   &  -    & -    \\
\  U             &  -   &  -    &   -  &  -     &  -     &  -    &  -   &   -   &  .01   & .01 &   .01 &   .10\\
\  t (Myr)       &  -   &  -    &   -  &  -     &  -     &  -    &  -   &   -   &  0.0   & 0.0 &  4.50 & -     \\
\  \Ts (10$^4$ K)         &  -   &  -    &   -  &  -     &  -     &  -    &  -   &   -   &   -    & -   &  -    &  5\\
\  w(MC1)       &90    &190     & 5    &  -     &  -     &5      &  -   &20     &10      & -   &  1    &  -   \\
\  w(MC2)     &9000    &  -     &  -   &500     &  -     &  -   &50     &    -  &   -    & -   &   -   &  1   \\
\ &&&&&&&&&&&&&\\
\ Nuclear region (MC1) :  && &&&&&&&&&&&\\
\ &&&&&&&&&&&&&\\
\ \% [OII]  3727+    &   3.0 &  23.5  &  1.0  &   .0 &    .0 &    .0  &   .0 &  62.0   & 4.1 &    .0 &   6.3    & .0\\
\ \% [OIII]  4363    &   3.5  & 78.0   & 4.2  &   .0   &  .0   &  .0   &  .0 &  11.0   &  .6   &  .0 &   2.0   &  .0\\
\ \% [OIII]   5007+  &     1.2 &  25.0 &   1.6 &    .0 &    .0 &    .0  &   .0&   59.3 &   3.6  &   .0 &   9.2 &    .0\\
\ \% [OI]   6300+    &    .0 &   8.6 &   5.2 &    .0  &   .0   &  .17 &    .0&   57.4 &   4.6  &   .0 &  24.0  &   .0\\
\ \%  \Hb           &  5.3  & 10.3  &  2.4   &  .0   &  .0   &  .1    & .0 &  65.3  & 11.6  &   .0  &  4.9   &  .0\\

\ &&&&&&&&&&&&&\\
\ Regions  A and B (MC2) :&&&&&&&&&&&&&&\\
\ &&&&&&&&&&&&&\\
\ \% [OII]   3727+  &   68.7  &  .0  &   .0 &  26.9   &  .0  &   .0  &  2.8   &  .0  &   .0  &   .0 &    .0 &   1.6\\
\ \% [OIII]   4363  &    81.5  &  .0 &    .0 &   5.6  &   .0  &   .0  &   .6  &   .0  &   .0  &   .0 &    .0 &  12.4\\
\ \% [OIII]   5007+ &    61.5  &  .0 &    .0 &  27.7  &   .0  &   .0  &  3.1  &   .0  &   .0  &   .0 &    .0 &   7.6\\
\ \% [OI]   6300+   &   2.4   & .0   &  .0  & 78.5   &  .0   &  .0  &  1.65  &   .0   &  .0  &   .0  &   .0 &  17.4\\
\ \% \Hb &   65.40   &  .0  &  .0 &  27.0   &  .0   &  .0  &  1.3   &  .0   &  .0   &  .0  &   .0  &  6.3\\

\hline

\end{tabular}

\flushleft

$^1$ SD=shock dominated; pl=power-law; sb=starburst; hII: HII region

$^2$ in 10$^{-3}$ \erg

$^3$ in photons cm$^{-2}$ s$^{-1}$ eV$^{-1}$ at 1 Ryd

\end{table*}

\subsection{Single-cloud models : the physical conditions of the emitting gas}

 Single-cloud  models  are shown in Table 1.
Some models that do not contribute to the
best fit are also included
 (e.g. M5 and M10) in order to understand
the  dependence of  the line ratios    on the different input parameters.

Interestingly, the models which  roughly fit the line ratios belong to all of   the different 
types  which are considered in the interpretation of  LLAGN spectra 
(Contini 1997, Contini et al.  2003) : 
SD models with a  relatively low (\Vs=100 \kms in M1) and high (\Vs=500 \kms in M3) 
shock velocities,
RD composite models corresponding to a relatively low \Vs ~and a 
low radiation flux from
the AC reaching the cloud (M4, M5, M6, M7, and M8),  composite models relative to a young 
starburst (M9 and M10) and to a starburst with an age of 4.5 Myr (M11), and 
a composite model corresponding to an HII region (M12).
Both radiation-bound and matter-bound models appear in  Table 1.

 Cid Fernandez et al (2004) and Gonzalez-Delgado et al. (2004)   claim that
the relative importance between collisional phenomena and the  photoionizing
radiation  which originates from the AC and/or from  starbursts, or even from HII regions,
can be deduced from the line ratios.
Very recently,  analysing new ground observations of LLAGN
and data by HST, 
 respectively, they distinguished between objects with high [OI] 6300/\Ha ~ ($>$0.25)
and low [OI] 6300/\Ha  ~relatively to the stellar population age and to the
presence of an active nucleus.
The data collected from the grids are given in Fig. 1 . It can be seen that
 the data referring to starburst models show [OI]/\Hb $<$ 0.75 ([OI]/\Ha $<$ 0.25, adopting \Ha/\Hb =3),
out of  few models calculated by \Vs=100 \kms, U=0.01,  an age  of t= 3.3-4.5 Myr, and D=3 pc.
In other words, these clouds  are more likely located in the surroundings of the starburst, and they are
 merging with the ISM.
All the other models showing [OI]/\Hb $\geq$ 0.75 refer to composite models with a power-law flux,
shock dominated models, and models for HII regions. 

The spectra calculated by   SD  models with \Vs=300 \kms (M2) and \Vs=500 \kms (M3)
and a preshock density  \n0 =300 \cm3,  are chosen  following our previous results that 
the velocity gradient increasing towards the AC is accompanied by a density gradient,
 and  because [NI]/\Hb ~are relatively high and  [SIII]/\Hb   ~relatively low.
Model M4  accounts for a relatively low ionization flux from the AC that
is  appropriate to LINERs.
The  composite model which accounts for a starburst  at early age and a small ionization parameter (M9)
shows a relatively  low \Vs.
As was pointed out by Viegas, Contini \& Contini (1999) and confirmed by Contini \& Contini
(2003) low shock velocities
are characteristic of young starbursts, which had not yet the time to develop
supernovae. The only model corresponding to an older starburst   (M11)
which could be considered corresponds to a low \Vs.
Most of the line ratios to \Hb ~are higher than observed, out of [OIII] 4363 and [NI] 5200+. 
Finally,  the  spectrum emitted from gas  illuminated by  a  HII region corresponding 
to  a temperature of 5 10$^4$ K and U=0.1 (M12)  
is characterised by low [OII]/\Hb ~line ratios. 

The single-cloud models fit the data very roughly. Yet they
can lead to a  better fit when summed up adopting relative weights. 

\subsection{Multi-cloud models : comparison with the data}

\begin{table}
\centering
\caption{Comparison of calculated with observed line ratios (\Hb=1)}
\begin{tabular}{lllllllllll}\\ \hline
\     line   &obs$^1$ &MC1&obs$^2$&obs$^3$ & MC2   \\
\hline
\ [OII]3727+ &8. $\pm$ 0.7&7.4& 3.6 & 4.5 &4.  \\
\ [NeIII]3868&0.7$\pm$ 0.1&1. &- & -&-\\
\ [OIII]4363 &0.7$\pm$ 0.1 &0.2&-&-  &-& \\
\ [OIII]5007+&7.1 $\pm$ 0.6&7. &  1.& 1.3 &1.6  \\
\ [NI]5199+    &0.79$\pm$ 0.09 &0.07& -& 0.36&0.06  \\
\ [OI]6300+    &1.46$\pm$ 0.14&1.66&0.27& 0.32&0.7\\
\ [NII]6548+  &7.9 $\pm$ 0.7 &5.4& 7.8 & 9. &2. \\
\ \Ha ~n       &3.2&3.&3. & 3. &3.   \\
\ \Ha ~b       & -&-&-&-&-   \\
\ [SII]6716  & 1.8$\pm$ 0.2&1.6& 2.25& 2.49&1.3 \\
\ [SII]6730  &1.8$\pm$ 0.2&2.2& 1.89&1.86&1.1 \\
\ [OII]7325+ &  0.61 & 0.8&-&-&-\\
\ [SIII]6069+ & 0.29$\pm$ 0.05&1.8&-&-&-\\ \hline

\end{tabular}

\flushleft

$^1$ Gonzalez-Delgado \& Perez (1996), nuc. (1.5"), corrected with C(\Hb)=0.97

$^2$ Gonzalez-Delgado \& Perez (1996), A ($\leq$ 20"), not corrected

$^3$ Gonzalez-Delgado \& Perez (1996), B ($\leq$ 20"), not corrected

\end{table}

Modelling  the NLR of  different galaxies (Seyfert types 1 and 2, NLS1, 
luminous infrared starburst galaxies,
LINERs, etc) we have learned that the spectra are generally explained by multi-cloud models.
Models MC1 and  MC2 which appear in Table 2  result from the weighted averages
of single-cloud models  which better fit the line ratios. 
The weights corresponding to the different single-cloud models are given in the 
 rows  13 and 14 from the bottom of Table 1. 
The relative weights compensate for the \Hb ~absolute flux intensities
and indicate roughly the  covering factors of each type of clouds.
So e.g. the relative weight of model M12 is low, but this does not mean that 
HII regions  do not contribute  somewhat  to the NGC 4579 spectra.

In Table 2 we compare the observed  line ratios with model results.
The results show that  the N/H relative abundance is higher than cosmic (9.1 10$^{-5}$)
by a factor $<$ 2 within 1.5" and by a factor $>$ 3 in the extended region.
The [OIII] 4363/5007 line ratios are underpredicted by
a factor of 3, while [NI]/\Hb ~and [SIII]/\Hb ~are underpredicted and overpredicted
by factors of $\sim$ 6, respectively in the nuclear region. The fit, in fact could be improved by
models calculated  purposely for NGC 4579 and/or by a larger number of single-cloud models.
The average spectra corresponding to the extended regions A and B (MC2)  overpredict the
[OI]/\Hb ~line ratio and underpredict the [SII]/\Hb ~ratios. Generally, a lower [OI] flux
corresponds to a lower [SII] flux, because the first ionization potential of S is lower
than that of H and, therefore,  of O. So, we conclude that  also the S/H relative abundance
should be about twice the cosmic one (1.6 10$^{-5}$) in the extended region.
The number of  the observed line ratios is small
and does not constrain strongly  the averaged model.  Nevertheless,  the present
analysis, which is carried  out only  by  models  from the grids, shows  that the choice of  models 
presented in the grids  is sensible. 

The results show that in NGC 4579  the AGN coexists with  starbursts
and HII regions, as is generally found for LLAGNs.
The relative importance of the  different types is determined by the relative weights
adopted summing up the single-cloud models.

In the bottom of Table 1
the  contribution in percent of each model to each of some significant  lines  
([OII] 3727+, [OIII] 4363, [OIII] 5007+,  [OI] 6360+, and \Hb) 
is given for the nuclear region and in regions A and B. 

It can be noticed that the largest contribution ($\sim$ 60 \%) to the  nuclear spectra 
 comes from RD clouds corresponding to the highest flux (log \Fh = 10) and relatively
low velocities (M8). SD clouds corresponding to \Vs=300 \kms (M2) also contribute by factors
$>$ 8.6\%,  with maxima of $\sim$ 80 \% to the [OIII] 4363 line. 
 A contribution $<$  5\% to all the lines
comes from SD clouds with \Vs=500 \kms (M3).
Interestingly, a young starburst  (M9) contributes  by less than 13 \%
with maxima   of $\sim$ 12 \% to  [SII] 6717   and   \Hb.
Moreover, a starburst with an age of 4.5 Myr (M11) contributes to all the lines.

Shock velocities of 100 \kms  in the young starburst are reasonable, because the 
starburst had not yet the time
to develop SN. However, the low \Vs ~(100 \kms) and \n0 (100 \cm3) in the older starburst  
may indicate that the starburst
is in the outskirts of  nuclear region where the outward velocities are generally low, while
 the young starburst is
most probably correlated  with the jet within the nuclear region.
A similar situation was found in the Seyfert 2 galaxy NGC 7130 (Contini et al. 2002)
where a young starburst was revealed by modelling the spectra in the nuclear region,
even if the effect of the AGN dominates, while the circumnuclear regions show spectra
 corresponding to stars with lower temperatures.

Regions A and B show a strong contribution from SD clouds  (M1) and RD clouds (M4,  M7)
reached by a rather low flux,    and having \Vs=100 \kms.
HII regions corresponding to a temperature of 5 10$^4$ K (M12) are also present. 

It is clear that the HII regions are located in the circumnuclear regions of NGC 4579, while
the AGN and the young starburst coexist in the nuclear region.
The  distances of regions A and B from the nucleus are within  1 kpc 
(Gonzales-Delgado \& Perez 1996, Fig. 4a). The modelling shows that
 high \Vs ~do not appear in these regions.

Indeed, the classification  (Cid  Fernandes et al, Gonzalez-Delgado et al 2004)
between LINERs and Transition Objects considering the [OI]/\Ha ~line ratio
sets NGC 4579  among LINERs and is consistent with the presence
of an AGN nucleus  and of an old stellar population.

Finally,
the  spectra observed within  4.7" and 4.0" by
 Keel (1983) and Stauffer (1982), respectively,    
  hint about the conditions beyond the nuclear region.
The spectra show  significant contributions from HII regions  (M12), SD clouds with \Vs=500 \kms (M3),
and of  an AGN with  log \Fh=9 (M4).
It is  typical of the  NGC 4579 continuum SED that
the old star population strongly dominates in the optical range (see Sect. 3). These stars
 do not  affect the spectra
because their temperatures are too low to heat and ionize the surrounding gas.
The signature of old stars in the spectra could  be revealed by collisional events
in supernova winds leading to  high velocity shocks which
 contribute to the heating and ionization of the gas in the clouds.

\section{The continuum SED}

\begin{figure*}
\includegraphics[width=88mm]{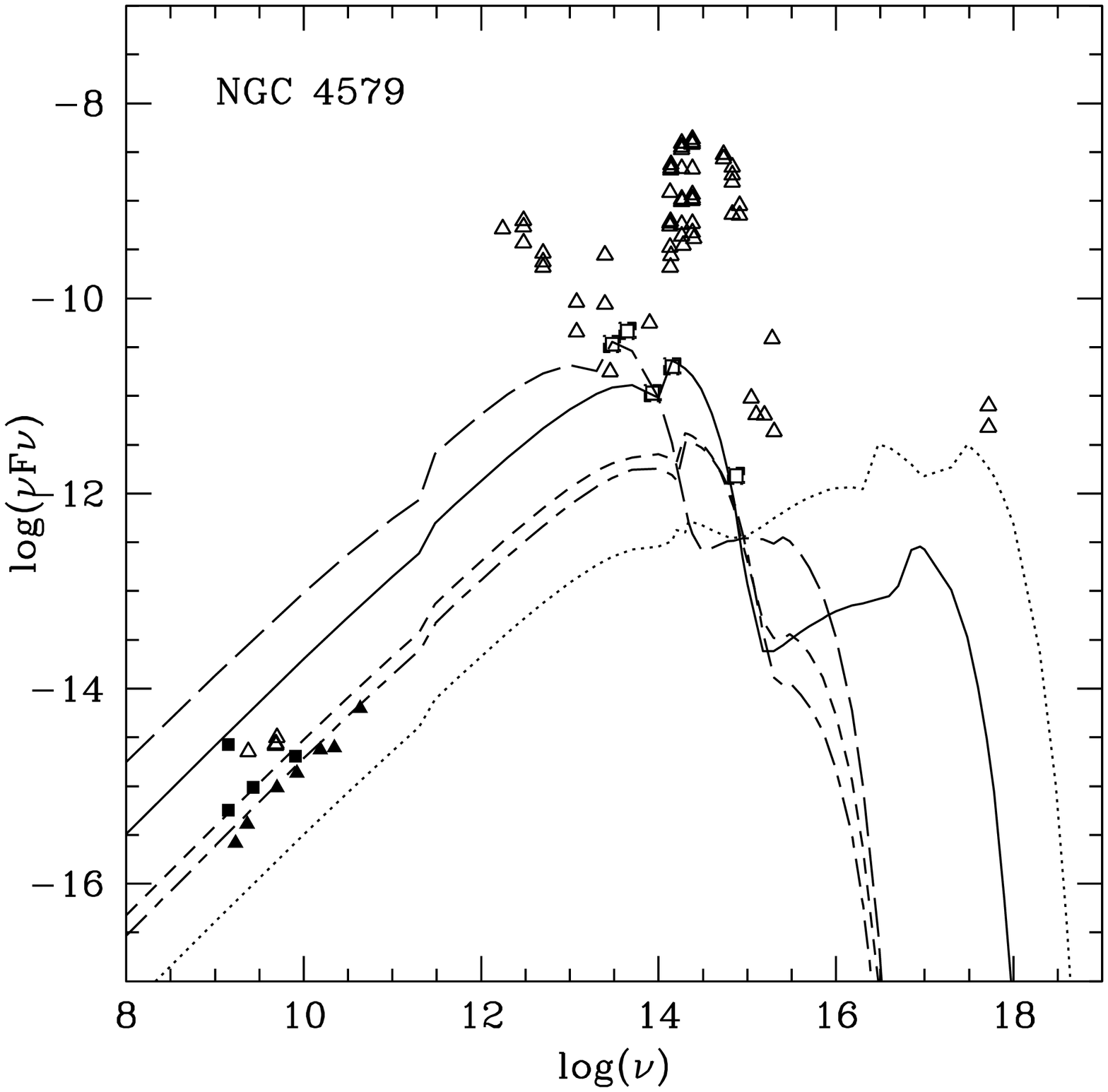}
\includegraphics[width=88mm]{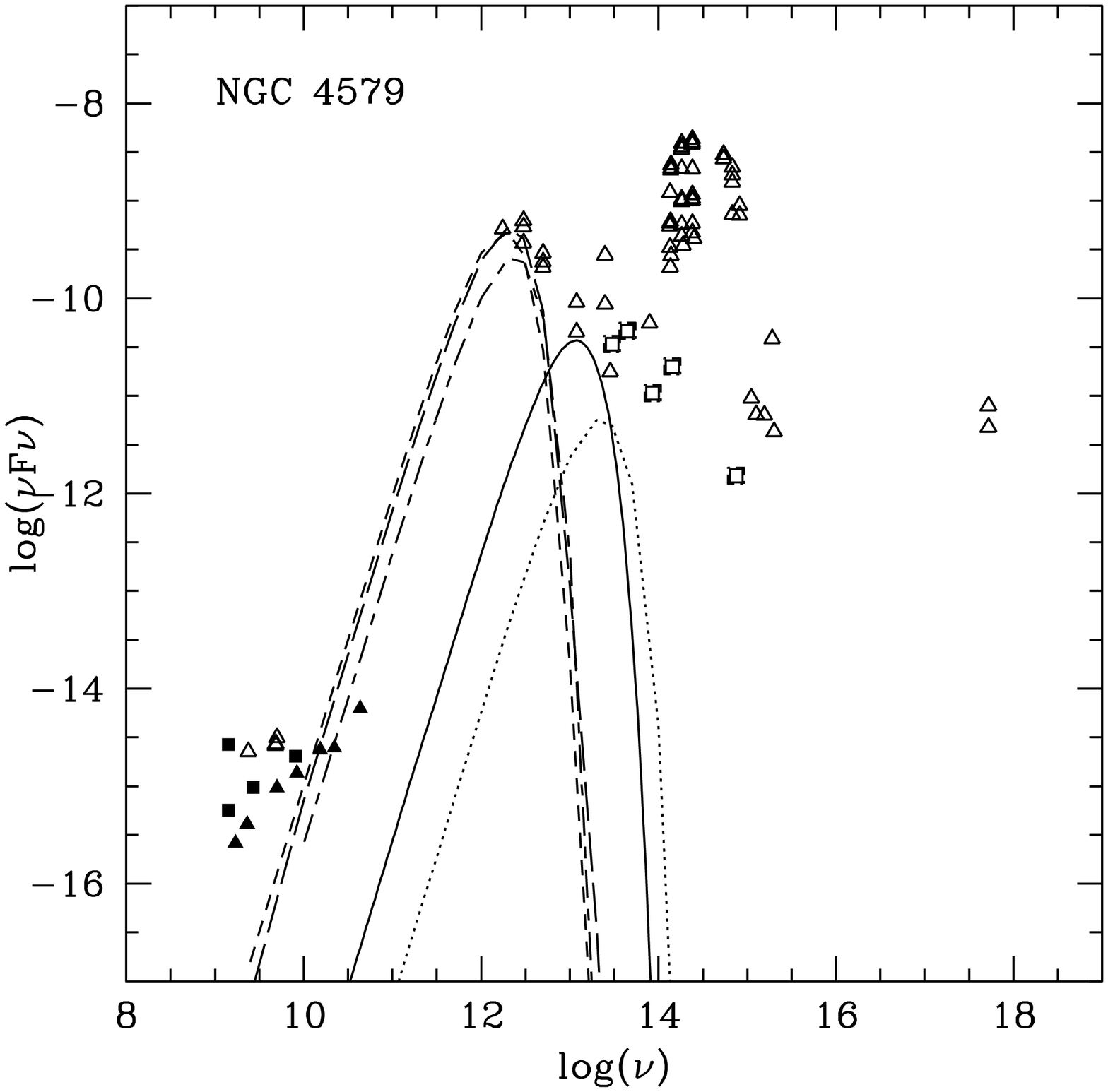}
\includegraphics[width=88mm]{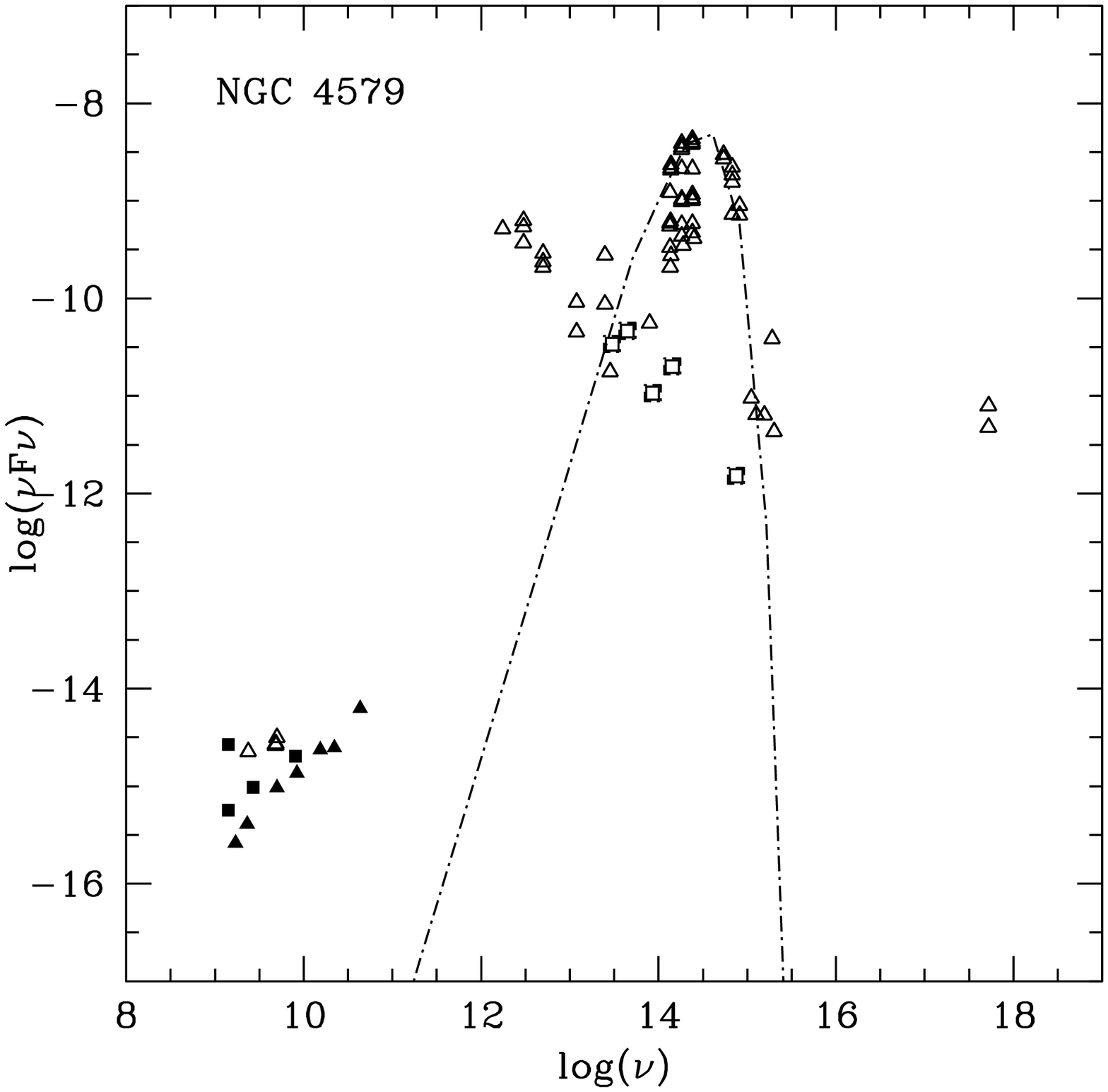}
\includegraphics[width=88mm]{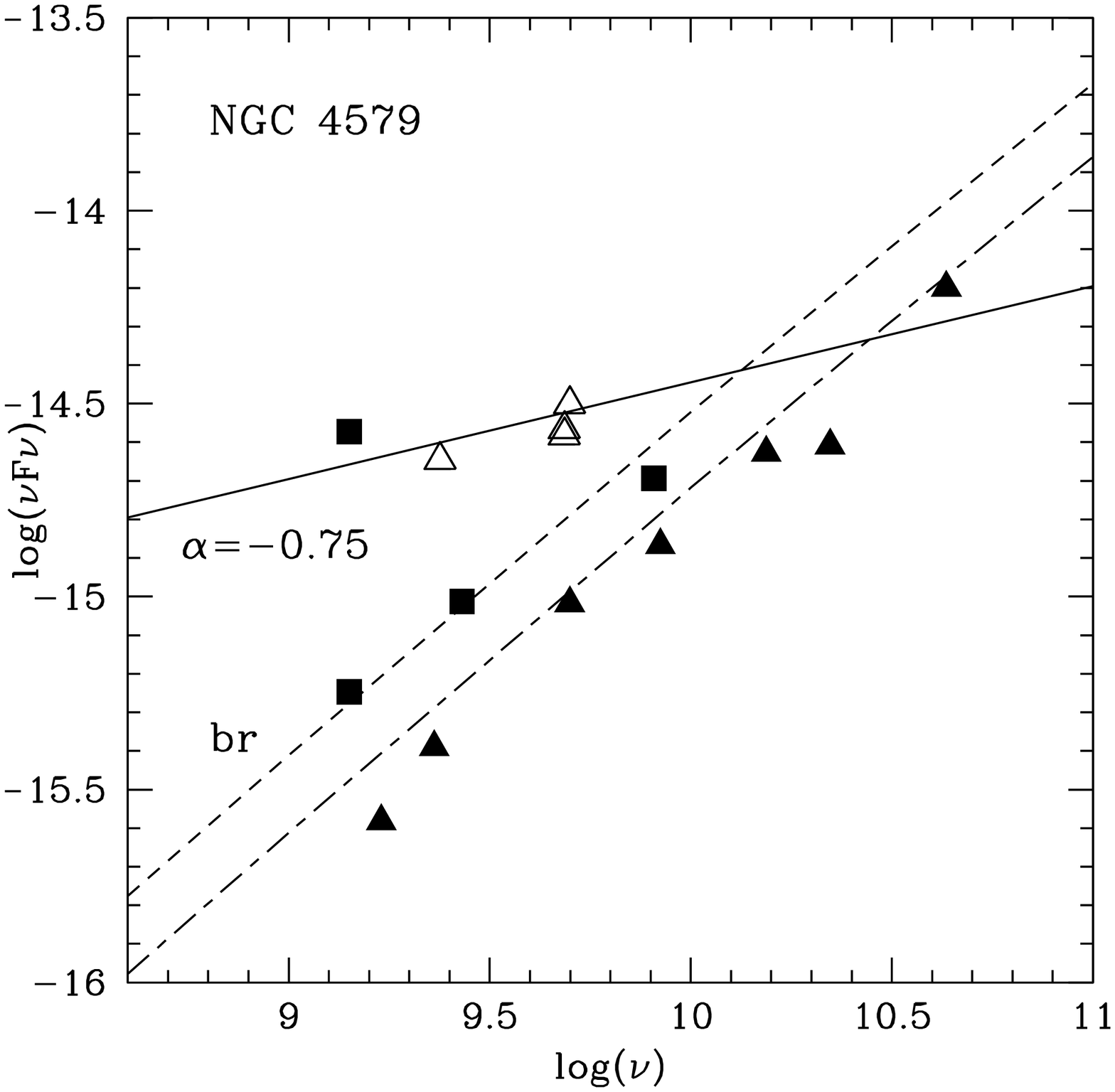}
\caption
    {The SED of the continuum.
Top-left panel: the bremsstrahlung corresponding to the different models.
 M4 (short-dashed lines), M3 (solid lines), M8 (short-dash-long-dashed lines)
  model corresponding to \Vs=1000 \kms
(dotted lines), model with \Vs=100 \kms, \n0=100 \cm3 (long-dashed lines);
top-right : dust emission (same notation as for top-left); bottom left:
black-body radiation from the old population stars; bottom right :
the slopes in the radio range: synchrotron radiation from the Fermi mechanism
(solid line), bremsstrahlung and dust emission from models M4 and  M8.
}

\end{figure*}

The  modelling  of line and continuum spectra must be cross-checked
until  the best fitting model is found  for both.

The bremsstrahlung emitted from  gas in the downstream region 
of the NLR clouds, generally
explains the observed SED in a  wide frequency  range (10$^8$ - 10$^{18}$ Hz).
In the infrared range, however,  dust emission  dominates, while
 the  old stellar population background is  observed 
in the optical - near-IR.
Synchrotron radiation  generated by the Fermi mechanism at the shock front
is often observed in the radio range of AGN.

In Figs. 2   we present the SED of the continuum of NGC 4579 in the different spectral ranges,
while  Fig. 3 shows the SED of the multi-cloud spectrum in a large frequency range.
Open triangles represent data from the NED/IPAC Extragalactic Database)
(Fabbiano et al. 1992, Kinney et al. 1993, Kodaira et al. 1990, De Voucouleurs et al. 1991,
Zwicky et al. 1961, Spinoglio et al. 1995, Aaronson 1977, Jarrett et al. 2003, Roussel et al. 2001,
Scoville et al. 1983, Soifer et al. 1989, Moshir et al. 1990, Tuffs et al. 2002,
Whiteoak 1970, Becker et al. 1991, Gregory \& Condon, 1991, Dressel \& Condon 1978,
Heeschen \& Wade 1964),
 where the data in the soft X-ray range
are upper-limit. Open squares  show the data from observations in the near-mid-IR by 
Alonso-Herrero et al (2003).  Black triangles represent the data of Anderson, Ulvestad, \& Ho (2003)
from 1.7 to 43 GHz using the Very Long Baseline Array, while the black squares
show the data by Hummel (1980) obtained at  the Westerbork Synthesis Radio Telescope, 
and the data by Crane (1977).
Notice that Hummel presents two data at 1.5 GHz, namely, the flux density  of the central source  
($<$ 21", i.e. $<$ 1.6 kpc) and the total flux density of the galaxy 
with a radio extent  of 3.9 arcmin along the major axis.
 The fluxes  from Crane 1977 at 2.7 and 8.1 GHz with size $<$ 3" (about 230 pc, adopting  a distance
to Earth of 16 Mpc as given by Roussel et al 2001) are also  given by Hummel (1980).
The data from the  NED   by Heeschen \& Wade (1969) 
  broad band measurement, as well as
 the total flux  at 1.5 GHz  by Hummel account for radiation from a more  extended region.

\subsection{Bremsstrahlung}

Consistent modelling of line (Sect. 2) and continuum spectra (Contini, Viegas \& Prieto 2004)
of NGC 4579
shows that the SED is explained by a multi-cloud model which accounts for SD
clouds  with shock velocities of  500 \kms (solid lines)
and RD clouds with a
low \Vs (100 \kms)  reached by a relatively low photoionization flux from the AC (log \Fh=9)
represented by short-dashed lines in Fig. 2. The  bremsstrahlung from these RD and SD clouds
(Fig. 2, top-left) fits the data in the optical-UV range.
The modelling of line spectra in the previous section revealed that the dominant component
from the nuclear region corresponds to RD clouds with \Vs=100 \kms, \n0=300 \cm3,
and log \Fh = 10 (short-dash-long-dashed lines).

A small contribution from high velocity clouds with \Vs = 1000 \kms and a preshock
density \n0=1000 \cm3 can explain
the emission in the soft X-ray range (dotted lines). This model however
 does not contribute to the line ratios because  the line fluxes are low.
Moreover, it was found by modelling the mid-IR data (Contini et al 2004)  from the sample 
of Seyfert galaxies observed by Alonso-Herrero et al. (2003)  that
a  high density, low velocity  single-cloud model (\n0=1000 \cm3, \Vs=100 \kms)
contributes to the fit of most of the galaxies, among them also NGC 4579.
This model which is also shown in Fig. 2 (long-dashed lines) corresponds to very low line 
ratios to \Hb ~and is not included in the selected grid (Table 1).

Recall that   the modelling of the  spectra  have to 
account for covering factors, for the distance   of the emitting clouds from the AC and 
for the distance of the galaxy  from Earth,
because the models are calculated at the nebula, while   the spectra are observed at Earth.
The models in Fig. 2 are  scaled roughly by taking into account  the weights 
and the above mentioned factors. In fact,  same types of clouds are located in different regions.
Notice that the model corresponding to the HII regions does not  contribute to the continuum because 
its  relative weight is low. 

\subsection{Infrared bump}

Dust and gas  are  coupled across  the shock front  and downstream,  so the grains are
collisionally heated by the gas to temperatures which depend on the shock velocity.
Reradiation from dust  in the IR is well explained in Fig. 2 (top-right)  by the same models which  
fit the bremsstrahlung.  
For all models a dust-to-gas ratio of 10$^{-14}$ by number (corresponding to 4 10$^{-4}$ by
mass adopting silicate grains) leads to a good fit of the infrared emission bump.
In composite models corresponding to a relatively low flux from the AC
or to  a low ionization parameter, U, dust grains are mainly heated collisionally by the gas
throughout the shock front and downstream. Therefore, the frequencies corresponding to the
peaks of the infrared emission depend on the shock velocity (see Contini et al. 2004).
Thus, it is difficult to distinguish between starburst activity and/or an active nucleus
from the analysis of the infrared bump.

\subsection{Old star contribution}

Fig. 2 (bottom-left panel) shows that  a   peculiar characteristic of NGC 4579  consists in the very high 
contribution to the continuum
in the optical range of  an old star population background, which is modelled by a black-body
corresponding to T=5000 K (dot-dashed line). 
As was explained by Contini et al. (2004)
Alonso-Herrero et al. data  (open squares)    are  uncontaminated  from the star contribution.

\subsection{Radio emissions}

 The data in the radio range, relative to different apertures, are used  to determine the different
slopes of the continuum  which correspond to the different mechanisms.
 Fig. 2 (bottom-right panel) shows a detailed presentation of the data  and modelling.  

Bremsstrahlung emission from high density gas downstream can be self absorbed
at long wavelengths,
as was found in the Circinus galaxy, NGC 7130, etc. This  occurs  to the downstream emission  from clouds 
with a preshock density \n0 $\geq$ 300 \cm3 and \n0=1000 \cm3.  
For example,  model M3 with \Vs=500 \kms has
a density downstream of 10$^4$ \cm3, a temperature of 10$^4$ K after recombination in a region
of about 10$^{19}$ cm,  leading to optically thick  gas ($\tau$ $\geq$ 1) for $\nu$ $\leq 10^{10}$ Hz.

On the other hand, in RD clouds, the bremsstrahlung from gas downstream  of the low velocity shock
explains the data from Hummel (1980) and Crane (1977) which refer to a region
extended to intermediate distances from the AC.
This is in agreement with Stauffer (1982) who claims that in his sample, nuclear 
radio sources and optical activity are well correlated.

The data  relative to the most  extended region of the galaxy  show relatively high fluxes 
and a different trend.
The data are well  explained by the synchrotron  power-law radiation
 with $\alpha$ = -0.75 (Bell 1978) (thin solid line) created at the shock front
by the Fermi mechanism.
In fact, the values corresponding to large apertures may include anything from SN and SNR 
in the galaxy as a whole.

The data  within 0.1 pc from the AC  are  investigated by Anderson et al. 
Although the slope defined by the data resembles bremsstrahlung emission with some absorption
at longer wavelengths,  Ulvestad  \& Ho  (2001) claim that the free-free interpretation
is untenable because the source is much more compact ($<$ 0.05 pc) than the dimension of
nuclear tori. 
Advection-dominated accretion flows
provide a good fit to the spectral slopes but the luminosities are too high  relative to the black hole
masses calculated from the diffuse motions. 
In this paper we adopt the modelling by Anderson et al. who
explain the observed spectral slope by jets (see Sect. 1)

\begin{figure}
\includegraphics[width=88mm]{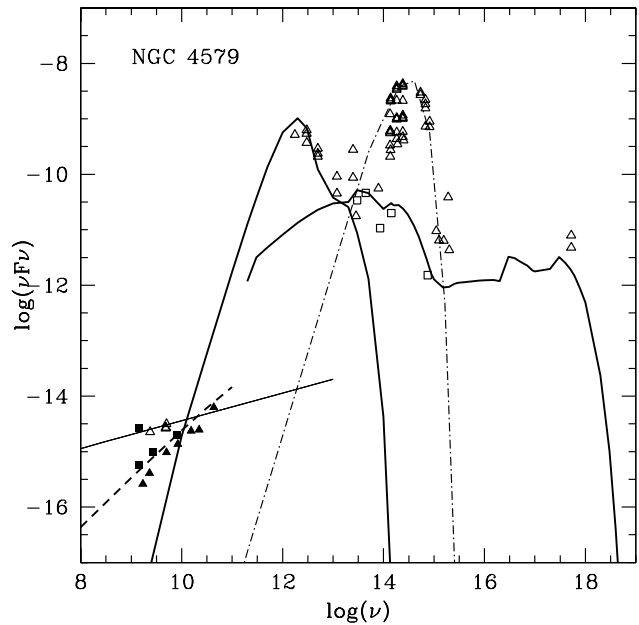}

\caption{The SED obtained by summing up the contribution of different models
in the different frequency ranges (solid lines). The dashed line shows that bremsstrahlung 
from high density clouds is self-absorbed.}

\end{figure}

\section{Discussion and concluding remarks}

\begin{figure}
\includegraphics[width=88mm]{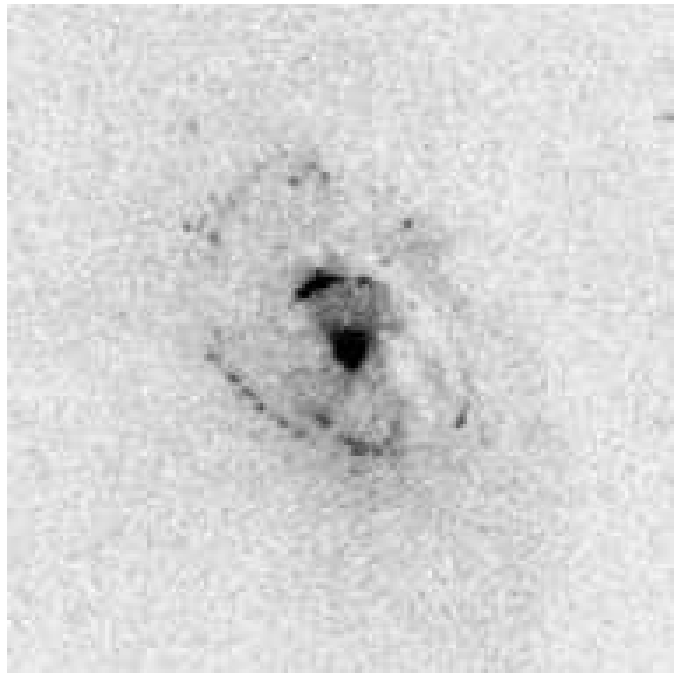}

\caption{
 $5"\times 5"$ section of a 
recent image of the nuclear region of NGC~4579,
obtained with the Hubble Space Telescope's Advanced Camera for Surveys
in the F330W band, corresponding to $\sim 3300$~\AA~
}

\end{figure}

In the previous sections we have modelled the line and continuum spectra of NGC 4579.

Early interpretations of the  line spectra  (e.g. Stauffer 1982)  based on the high electron temperature
derived from the low
[OIII] 5007/4363 ratio and on  the similarity with the spectra of SNR, already  suggested the presence of 
shocks.
However, LINER-like spectra are also shown by galaxies in which
 some filaments, arcs and extended features  reveal 
 vigorous recent bursts of star formation (Gonzalez-Delgado \& Perez 1996, Contini 1997).
The weak broad \Ha ~emission observed in NGC 4579  and  UV radiation similar to those of Seyfert galaxies
suggest  an AGN  nature of the spectra  but with stronger low-ionization lines (Barth et al. 2001).

The results of the present modelling  show  that   the multi-cloud models which
best explain the line ratios  account for all types of photoionizing sources,
as well as for the shocks.
The relative importance of the different  ionizing mechanisms
is determined by the relative weight adopted to sum up the different single-cloud models.

The nuclear emission is  dominated by the AGN
with  a  relative low flux from the AC  (log \Fh=10) compared with that found
in the NLR of Seyfert galaxies (log \Fh $\sim$ 11-13).
On top of this there  is the relative small contribution of a young starburst
in agreement with  Maoz et al (1995) who found two emission components in the nucleus of NGC 4579.
Shock dominated clouds with velocities of 100, 300, and 500 \kms,
 as well as   an older starburst with an age of 4.5 Myr, probably
located in the external nuclear region  were also revealed by modelling.
The good agreement of  model predictions with   the observations 
is evident in Fig. 4. 
Fig. 4 shows a $5"\times 5"$ section of a 
recent image of the nuclear region of NGC~4579,
obtained with the Hubble Space Telescope's Advanced Camera for Surveys
in the F330W band, corresponding to $\sim 3300$~\AA~
 (D. Maoz et al., in preparation). The extended source $0.6"$ from the
nucleus, previously
 noted by Maoz et al. (1995), is now resolved into a short arc
which forms the beginning of the $\sim 1"$ spiral arm that can be traced
for over one revolution around the nucleus. This region likely contributes
young stellar features to ground-based spectra   where it is included
in the aperture.   

In the extended A and B regions ($\sim$ 1 kpc),  RD clouds  reached by a  flux  weaker by
a factor of 10 than that in the nuclear region, due to dilution with distance,
contribute  to  most of the  line fluxes,
while  HII regions corresponding to \Ts= 5 10$^4$ K are also revealed.
The contribution of SD clouds with \Vs=100 \kms
to the [OII] 3727 and [OIII] 5007 lines is significant,
confirming that the velocity gradient decreases towards the
outskirts of the galaxy.
 Such low velocities
are generally present in LINERs and LLAGNs (Contini, Viegas, \& Campos 2003).

The regions  observed by Stauffer (1982) and Keel (1983) within intermediate apertures 
 are  explained by  the same AGN
corresponding to a   low flux,  by shocks with a rather
high \Vs ~(500 \kms) and by  the HII regions  which  dominate  the spectra.
The high velocity SD clouds  created by collisional events in SN winds may be related with
 the strong flux from the old stellar population in the optical - near-IR range
 revealed by the continuum SED of NGC 4579.

Indeed, the calculated  \Vs ~ throughout the galaxy are in the range of those observed in the NLR of AGN.
 Stauffer (1982) claims
that the observed FWHM are  280 \kms,
while Barth et al find a FWHM of about 520-570 \kms for the narrow \Ha, and for
[NII] and [SII] lines.
 Pogge et al (2000)  claim that the bulk of \Ha ~and [OIII] emission
comes from a clumpy complex  with an elliptical shape of major axis
of 2" (the emission region is between  tens and hundreds of parsecs).
So many different conditions coexist and spectra from clouds with  different \Vs
~can  indeed be present.
The same conclusions were found modelling the Seyfert 2 galaxies Circinus (Contini, Prieto, \& Viegas
1998),  NGC 7130 (Contini et al. 2002),  and  LINERs (Contini 1997). In  Seyfert 2 galaxies
 the AGN  dominates in the nuclear region, while in LINERs the  shock prevails.

The complex nature of LLAGN  as an AGN + starbursts is discussed very recently 
by Cid Fernandez et al. (2004) and Gonzalez-Delgado et al. (2004) in the light of
new observations.
They found that 
 few LLAGN have a detectable young ($<10^7$ yr) starburst component, 
indicating that massive stars do not contribute significantly
to the optical continuum. No features of W-R stars are present.
Intermediate age populations are very common  in LLAGN with relatively
low [OI]/\Ha ~emission ($\leq$ 0.25), but rare  for stronger [OI].
Moreover,
the dominant source of ionization in strong-[OI] LLAGN is likely an AGN which is 
consistent with the detection of broad Balmer lines in emission in a few cases
and a larger contribution of older stars in their stellar population.
Therefore, it seems  that
the strong [OI] ([OI]/\Ha $\geq$ 0.25) objects are  true LLAGN, with stellar
processes being insignificant.

Actually, NGC 4579 belongs to the strong [OI] type.
The models presented in Table 1 which were selected  by fitting the
NGC 4579 spectra show that only models M1 and M10 correspond to a very low
[OI]/\Ha.   M10 is a matter bound model, which does not contribute to
NGC 4579 spectra, and M1 is a SD model corresponding to low \Vs ~and \n0.
On the other hand, the young starburst corresponding to radiation bound model M9,
  is characterised by
 [OI]/\Ha ~= 0.22 (adopting \Ha/\Hb = 3)), while the  power-law dominated model (M4 in Table 1)
and the high velocity shock (M3) show  high [OI]/\Ha ~line ratios (0.66 and 1.2, respectively).
This confirms that in low [OI]/\Ha ~LLAGN a young  starburst dominates,
as found by Gonzalez-Delgado et al., while, starbursts of older ages (t$\geq$ 3.3 Myr)
show very high [OI]/\Ha ~as for AGN. This does not exclude that
in complex high [OI]/\Ha ~objects,   a young starburst can  be present, however, by a low 
weight, as found for NGC 4579.
Also SD  models  show high [OI]/\Ha ~line ratio.
These models  generally explain the emission in supernova remnant, i.e
they  are related with an old stellar population, in agreement with
 Cid Fernandez et al who claim that
in the strong-[OI] LLAGN, stars cannot play an important role in the gas ionization.
Recall that  the presence of old stars  is revealed by  the SED of NGC 4579.

We have particularly focused on the  nature of the emission in the radio range. We have adopted
 the Ulvestad \& Ho explaination that    
synchrotron emission from the base of the jet fits the radio  data observed within 0.1 pc
from the active centre.
Modelling the continuum SED  it is found that radio emission  within intermediate distances (10-200 pc) 
from the centre is explained by the bremsstrahlung
from gas downstream of low velocity  shocks (\Vs=100 \kms) reached by a  rather low
radiation intensity from the AC. The shocks are
created by the collision of clouds with the jet (Blandford \& K\"{o}nigl 1979).
Within  larger distances from the AC ($>$ 100 pc) the dominant radio emission is synchrotron
radiation with a spectral index of  -0.75 created by the Fermi mechanism at the
shock front,  as is often  observed from Seyfert galaxies
(Contini \& Viegas 2000).
So all types of emission observed within  different  radius from the AC
can be  reconciled  consistently with the presence of the  jet
outflowing from the accretion disk, if shocks  are created by collision of the  jet
 with ISM and/or NLR matter.
Consistently with the presence of the jet, the young starburst can be created in the
nuclear region close to the base of the jet, by the interaction of the high velocity
gas with  clouds, while the old stars are  located in the circumnuclear region.
The stars were created by triggering in  collision  of shocks with dense matter, so 
the shocks are loosing velocity  towards the outskirts of the galaxy.

In conclusion, we have explained both the line and continuum spectra in the different
regions of the LLAGN NGC 4579  by composite models accounting for the AGN and for the shocks
which  are created by collision of the jets
with ISM  clouds  within different distances from the centre, as well as for
starbursts of different ages, and for the HII regions.
 
\section*{Acknowledgments}

I am very grateful  to  an anonymous referee for  valuable comments which improved the
presentation of the paper. I thank   D. Maoz for providing the  recent image of the nuclear region
of NGC 4579,
   S.M. Viegas  and M.A. Prieto for helpful discussions,
and  I. Goldman for reading the manuscript.

\bsp

\label{lastpage}

\end{document}